\documentclass[a4paper]{jjapproc}
\usepackage{jwps}

\usepackage{bm}

\title{A two-stage data-analysis method for total-reflection high-energy positron diffraction (TRHEPD)}

\author{Kazuyuki Tanaka${}^1$, Izumi Mochizuki${}^2$, Takashi Hanada${}^3$, Ayahiko Ichimiya${}^2$, Toshio Hyodo${}^2$, and Takeo Hoshi${}^1$}

\inst{${}^1$Department of Applied Mathematics and Physics, Tottori University, 4-101 Koyama-Minami, Tottori 680-8550, Japan  \\ 
${}^2$ Institute of Materials Structure Science, High Energy Accelerator Research Organization, 1-1 Oho, Tsukuba, Ibaraki 305-0801, Japan  \\ 
${}^3$Institute for Materials Research, Tohoku University, 2-1-1 Katahira, Aoba-ku, Sendai 980-8577, Japan}

\email{hoshi@tottori-u.ac.jp}

\recdate{October XX, 2020}

\abst{Total-reflection high-energy positron diffraction (TRHEPD) is a novel experimental method for the determination of surface structure, which has been extensively developed at the Slow Positron Facility, Institute of Materials Structure Science, High Energy Accelerator Research Organization (KEK). 
In this paper, a two-stage data-analysis method is proposed.
The data analysis is based on an inverse problem in which the atomic positions of a surface structure are determined from the experimental diffraction data (rocking curves). The relevant forward problem is solved by the numerical solution of the partial differential equation for quantum scattering of the positron. In the present two-stage method, the first stage is a grid-based global search and the second stage is a local search for the unique candidate for the atomic arrangement. The numerical problem is solved on a supercomputer.
}

\begin{document}

\maketitle





\section{Introduction}

Fast, reliable computational methods used in the analysis of observed data can lead to innovative improvements in experimental measurement techniques. An example is found in the development of cryo-electron microscopy, which won the Nobel Prize in Chemistry 2017.

The present paper focuses on a computational analysis method for total-reflection high-energy positron diffraction (TRHEPD), a novel measurement technique for atomic scale surface structure. At the Slow Positron Facility (SPF), Institute of Materials Structure Science (IMSS), High Energy Accelerator Research Organization (KEK), much work has been conducted to successfully reveal surface structures of interest (see reviews  ~\cite{FUKAYA2018-JPHYSD, FUKAYA2018-MONOATOMIC} and papers \cite{Mochizuki2016_TiO2, Fukaya2016_TRHEPD_Ge_on_Al, Endo2020_TRHEPD_Ca_SiC}). A previous paper \cite{TANAKA2020_ACTA_PHYS_POLO} reports the recent activity of the software development for TRHEPD data analysis. The software developed was based on a local search algorithm starting with an initial guess of the atom positions.

The present paper proposes, as a following advancement, a two-stage analysis method in which we first use a global search algorithm and then a local search algorithm for a final solution. The method does not require any initial guess, unlike the previous one \cite{TANAKA2020_ACTA_PHYS_POLO}. Since the computational time cost of a global search would be too large for a personal computer, the present software was developed to use a supercomputer in a modern massively parallel architecture.

The present paper is organized as follows: 
the experiment and theory of TRHEPD is explained briefly in Section ~\ref{SEC-TRHEPD}; 
Section ~\ref{SEC-DEMO} details a technical demonstration of the software and several discussions. Summary and future works are described in Section  ~\ref{SEC-SUMMARY}.

\begin{figure}[h]
\begin{center}
  \includegraphics[width=0.95\textwidth]{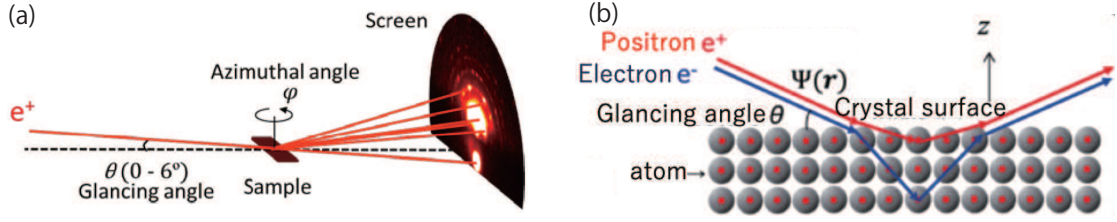}
\end{center}
\caption{Schematic figures of (a) experiment and (b) theory of TRHEPD. In (b), the diffracted beams are schematically shown not only for the positron beam in THREPD experiment but also for the electron beam in reflection high-energy electron diffraction (RHEED) experiment. The set-ups of the two experiments are essentially the same. 
}
\label{FIG-THREPD-OVERVIEW}       
\end{figure}

\section{METHOD \label{SEC-TRHEPD} }

The TRHEPD method, illustrated schematically in Figure \ref{FIG-THREPD-OVERVIEW}, will be briefly described in this section. The experimental set-up of TRHEPD is essentially the same as reflection high-energy electron diffraction (RHEED). The penetration depth of the positron is much smaller than that of the electron, as schematically shown in Figure \ref{FIG-THREPD-OVERVIEW}(b), since the electrostatic potential of in every material is positive. Therefore the positron probes selectively the top-most surface layer and only a few subsurface layers, and thus the positron is an ideal probe of surface structure.

The data-analysis problem of TRHEPD is to determine the positions of the atoms at the top-most surface layer and a few subsurface layers. Hereafter, the $z$ axis is chosen to be perpendicular to the sample surface and the position of the $i$-th atom is denoted as  $\bm{r}_i = (x_i, y_i ,z_i)$. 
We denote $N_{\rm a}$ as the number of the atoms whose positions 
$((x_i,y_i,z_i), i=1,2,...,N_{\rm a})$ are to be determined.

The data analysis is based on the inverse problem in which  
the surface structure $X$ is determined from the experimental diffraction data, 
$D_{\rm exp} (D_{\rm exp} \Rightarrow X)$.
The calculated diffraction data, $D_{\rm cal}$,
can be obtained for the assumed structure $X$;
this process  is called the forward problem $(X \Rightarrow D_{\rm cal}(X))$.

\subsection{Experiment of TRHEPD}


The TRHEPD experiment is shown schematically 
in Figure \ref{FIG-THREPD-OVERVIEW}(a). 
The incident wave direction is characterized
by the glancing angle $\theta$ and the azimuthal angle $\varphi$. 
We also define
the  incident wave vector $\bm{K}^{\rm (in)} = (K \cos \theta \cos \varphi, K\cos \theta \sin \varphi, 0)$
projected on the $x$-$y$ plane.

Diffraction spots on the screen in 
Figure \ref{FIG-THREPD-OVERVIEW}(a) are characterized by the two-dimensional indices $(p,q)$ of reciprocal lattice rods. The intensity of a particular spot, $D_{pq}$, is observed as a function of the incident glancing angle, $\theta$ ($D_{pq}=D_{pq}(\theta)$)  and called the rocking curve.

The rocking curve, $D_{pq}(\theta)$, depends also on the azimuthal angle ${\varphi}$,
which is fixed during a rocking curve measurement.  
The spot with the indices of $(p,q)=(0,0)$ is usually 
the brightest 
and called 00 spot. 
This paper focuses on the intensity of the 00 spot and 
we drop the indices for simplicity
($D=D(\theta)$). 
The observed data for discrete glancing angles is denoted
as $\bm{D} \equiv (D(\theta_1), D(\theta_2), ..., D(\theta_{\nu}))$,
where a typical number of the glancing angles is $\nu=$ 50 - 100. 
The present data analysis is carried out
on the normalized vector data $\bm{D}$ ($|\bm{D}|=1$) and similarly normalized calculated values.
TRHEPD patterns, and 
hence the rocking curves extracted from them,
are hardly affected by the atomic coordinate parallel to the vector $\bm{K}^{\rm (in)}$,
because the energy of the incident beam is typically 10 keV 
with wave length much shorter than the lattice constants of materials.
Thus if the vector $\bm{K}^{\rm (in)}$ is parallel to the $y$-axis, for example, 
the rocking curve depends only on the $x$ and $z$ components of the atomic position 
($\bm{D} = \bm{D}(x_1,x_2,...,x_{N_{\rm a}},z_1,z_2,...,z_{N_{\rm a}})$). 
One can choose the azimuthal angle 
intentionally shifted from any low-index zone axis so that 
practically the rocking curve depends almost only on the $z$ components of the atom position 
($\bm{D} = \bm{D}(z_1,z_2,...,z_{N_{\rm a}})$).
This is called the one-beam condition \cite{ICHIMIYA1983_OB}.
Other choices of the azimuthal angle are called the many-beam condition,
where the azimuthal angle is set so that the beam is incident along a low-index zone axis. 
In order to determine the atomic positions in the $x$ and $y$ directions independently, two different azimuthal angles in the many-beam condition are adopted sequentially.
Then the rocking curve depends only on the $x$ or $y$ component, since
the $z$ component already known by the analysis of the data in the one-beam condition could be fixed. 
These properties  allow us 
to reduce the number of variables in the  analysis of each data set from $3N_{\rm a}$ to $N_{\rm a}$.
Such dimensional reduction is of great advantage
in realizing fast and reliable data analysis.  

Therefore, the measurement procedure, typically, consists of three diffraction data sets for different azimuthal angles. 
One data set is for the one-beam condition, which is denoted as $\bm{D}^{\rm (OB)}$. 
The other two sets are for the many-beam condition, 
which are denoted as $\bm{D}^{\rm (MB1)}$ and $\bm{D}^{\rm (MB2)}$.
The incident wave vectors $\bm{K}^{\rm (in)}$ are orthogonal between the two data sets in the many-beam condition. 
The first procedure in data analysis determines the $z$ component of the atomic position 
$(z_1, z_2, ..., z_{N_{\rm a}})$ from the data set in the one-beam condition $(\bm{D}^{\rm (OB)})$. 
The second procedure with the already determined z components $(z_1, z_2, ..., z_{N_{\rm a}})$ fixed determines each component on the $x$-$y$ plane from each set of the two data sets in the many-beam condition $(\bm{D}^{\rm (MB1)},\bm{D}^{\rm (MB2)})$.

\subsection{Theory of TRHEPD}

The theory or the forward problem $(X \Rightarrow D_{\rm cal}(X))$
of TRHEPD \cite{ICHIMIYA1983} is based on 
the full-dynamical quantum scattering problem 
of the positron wavefunction $\Psi(\bm{r})$,
which is shown schematically in Figure \ref{FIG-THREPD-OVERVIEW}(b). 
The calculation method  \cite{ICHIMIYA1983} was originally developed for electron diffraction (RHEED) 
but is applicable here,
since the TRHEPD method is different from RHEED only in the sign of the charge of the incident particle. 
 As the electron beam penetrates into deeper layers than the positron, 
 owing to  refraction off the surface, as schematically shown in Figure \ref{FIG-THREPD-OVERVIEW}(b), 
electron diffraction has added complexities compared to positron diffraction.

The calculation method solves numerically
the partial differential equation with a given glancing angle and an azimuthal angle 
\begin{eqnarray}
\left( \Delta + U(\bm{r}) + E \right) \Psi(\bm{r}) =   0, 
\label{EQ-PDE}
\end{eqnarray}
so as to obtain the rocking curve data $\bm{D}=\bm{D}_{\rm cal}(X)$. 
Here $U(\bm{r})$ is the crystal potential determined
by the atomic positions $X$.
The crystal potential $U(\bm{r})$ is periodic on the $x$-$y$ plane and can be
written by the two-dimensional Fourier series
\begin{eqnarray}
U(x,y,z) =   \sum_{m} U_{m}(z) \exp \left( i (k_x^{(m)} x+ k_y^{(m)} y) \right),
\label{EQ-POT}
\end{eqnarray}
where $(k_x^{(m)} ,k_y^{(m)} )$ is the surface reciprocal lattice vector of the $m$-th rod.

The calculation under the one-beam condition \cite{ICHIMIYA1983_OB} is much faster
than that under the many-beam condition,
since only one Fourier component $(k_x^{(m)},k_y^{(m)})=(0,0)$ is  assumed to be non-zero
in Equation ~(\ref{EQ-POT}) 
\begin{eqnarray}
U(x,y,z) = U_{0}(z). 
\label{EQ-POT2}
\end{eqnarray}
under the one-beam condition.
Consequently, 
the wavefunction can also be written as $\Psi \equiv \Psi(z)$ and
Equation ~(\ref{EQ-PDE}) is reduced to 
a one-dimensional scattering problem 
\begin{eqnarray}
\left( \frac{d^2}{d z^2} + K^2 \sin^2 \theta + U_0(z)  \right) \Psi(z) =   0.
\label{EQ-ODE}
\end{eqnarray}
Several codes in Fortran have been developed 
to solve the forward problem
\cite{ICHIMIYA1983, ICHIMIYA1983_OB, Hanada1995_RHEED}
and we use the code in Reference ~\cite{Hanada1995_RHEED}
in the calculations of the present paper.

\begin{figure}[h]
\begin{center}
  \includegraphics[width=0.6\textwidth]{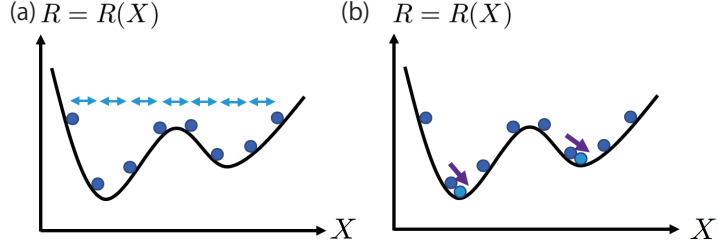}
\end{center}
\caption{Schematic figure of the present two-stage search algorithm
for finding the minimum point of a given function $R=R(X)$.
A grid-based global search is carried out 
in the first stage (a) and then
the local search is in the second stage (b).  
The initial points in the second stage are obtained in the first stage.
The function R(X) is called the reliability factor or R-factor.
}
\label{FIG-ALGORITHM}       
\end{figure}

\subsection{Data analysis algorithm \label{SEC-DATA-ANALYSIS-METHOD} }

This subsection details  
the two-stage method for the data analysis of the TRHEPD data. 
The method is based on 
the inverse problem  $(\bm{D}_{\rm exp} \Rightarrow X)$ and
the solution $X$ is determined by minimizing 
the residual function between
the calculated and experimental diffraction data
\begin{eqnarray}
R(X) \equiv | \bm{D}_{\rm cal}(X) - \bm{D}_{\rm exp}|
\label{EQ-R-FACTOR}
\end{eqnarray}
with a given experimental diffraction data $\bm{D}_{\rm exp}$.
The function $R(X)$ is called the reliability factor or R-factor. 

The two-stage analysis method in the present paper
is illustrated in Figure ~\ref{FIG-ALGORITHM}.
Figure ~\ref{FIG-ALGORITHM}(a) shows schematically
the first stage in a grid-based global search algorithm. 
A set of equi-interval grid points $\{ X^{\rm (g)}_1, X^{\rm (g)}_2, ..., X^{\rm (g)}_M \}$ is shown,
where $M$ is the number of the grid points.
Then the R-factor is calculated at each grid point, $\{ R(X^{\rm (g)}_1), R(X^{\rm (g)}_2), ..., R(X^{\rm (g)}_M) \}$, 
and local minimum points $X^{(0)}$ are found that give minimum R-factor values relative to the points around.
Since the calculations of the R-factor at the prepared grid points are independent,
the procedure is ideal for processing with a supercomputer
in the modern massively parallel architecture.  
Figure ~\ref{FIG-ALGORITHM}(b) shows schematically
the second stage in the local search algorithm.
The atomic position $X$ is updated so as to decrease $R(X)$ from the initial guess $X^{(0)}$,
until the position satisfies a convergence criteria
($X^{(0)} \Rightarrow X^{(1)} \Rightarrow X^{(2)}  .. \Rightarrow X^{(k)} \Rightarrow...\Rightarrow X^{\rm (sol)} $).
The local search algorithm employs
Nelder-Mead algorithm \cite{TANAKA2020_ACTA_PHYS_POLO, Nelder-Mead1965, Lagarias1998}.
We developed a Python-based data analysis code.
The parallelism in the global search is realized
by the Message Passing Interface (MPI) technique,
a standard technique for parallelism,
in the \verb|mpi4py| library (\verb|https://bitbucket.org/mpi4py/mpi4py|). 
The Nelder-Mead algorithm in the local search is realized
by the module in the \verb|scipy| library (\verb|scipy.optimize.fmin|). 
The method is standard and the use of the \verb|scipy| library is not essential.

\begin{figure}[h]
\begin{center}
  \includegraphics[width=0.9\textwidth]{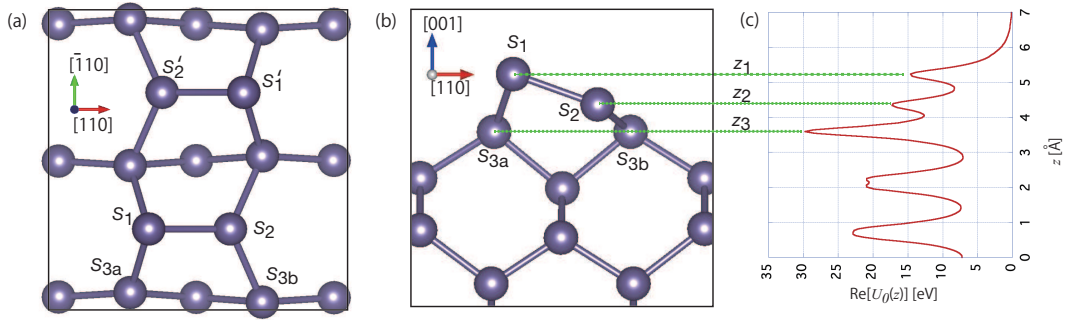}
\end{center}
\caption{
(a) The top view along the $[001]$ direction of the Ge(001)-c($4 \times 2$) surface
with asymmetric surface dimers of (S$_1$, S$_2$) and (S$'_1$, S$'_2$).
A surface dimer consists of 
the upper (vacuum side) atom denoted as S$_1$ or S$'_1$ and
the lower (bulk side) atom denoted as S$_2$ or S$'_2$.
The $z$ coordinates of S$_1$ and S$_2$ 
are denoted by $z_1, z_2$, respectively.
The atoms in the subsurface layer are denoted by S$_{\rm 3a}$ and S$_{\rm 3b}$
and their $z$ coordinate is denoted by $z_3$. 
(b) The side view along the $[\bar{1} 1 0]$ direction
for a part of the Ge(001) surface,
in which only the asymmetric surface dimer (S$_1$, S$_2$) and several subsurface layers are shown. 
(c) The real part of the scattering potential $({\rm Re}[U_0(z)])$ averaged over the $x$-$y$ plane 
is drawn for the structure in (a) and (b). 
}
\label{FIG-SURFACE}       
\end{figure}

\begin{figure}[h]
\begin{center}
  \includegraphics[width=0.7\textwidth]{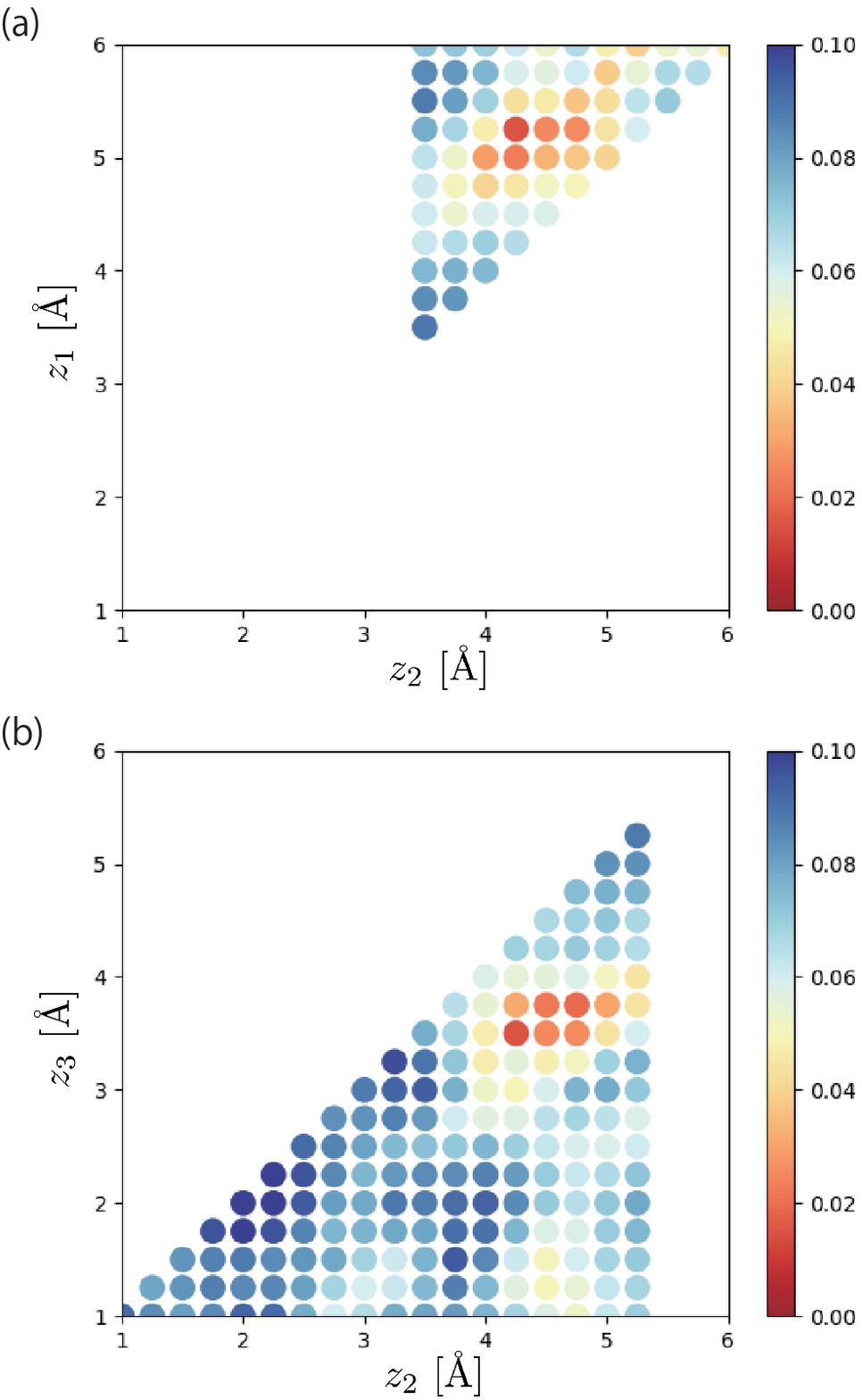}
\end{center}
\caption{
Partial regions of the global search. 
The calculated value of the R-factor $R=R(z_1, z_2, z_3)$ 
is plotted at each grid point as colored circles.
on (a) the plane of $z_3=3.50 \AA$
and (b) the plane of $z_1=5.25 \AA$. 
The grid points lie under the constraints 
of $1 \AA \le z_3 \le z_2 \le z_1  \le 6 \AA$.
}
\label{FIG-SEARCH-1}       
\end{figure}

\section{Test problem and the result \label{SEC-DEMO}}

\subsection{Test problem}
 
The two-stage algorithm was applied
in a numerical test problem in the one-beam condition
for the Ge(001)-c($4 \times 2$) surface structure. 
The structure is found in many papers, such as Reference \cite{STEKOL2002}
and the top view is shown in Figure \ref{FIG-SURFACE}(a).  
The first and second surface atoms 
are denoted by S$_1$ and S$_2$, respectively,
and form an asymmetric dimer.
The third surface atoms  
are denoted by S$_{\rm 3a}$ and S$_{\rm 3b}$.
A side view along the $[\bar{1} 1 0]$ direction
is shown in Figure \ref{FIG-SURFACE}(b).
The asymmetric surface dimers,
such as (S$_1$, S$_2$) and (S$'_1$, S$'_2$),
form a row in the $[\bar{1}10]$ direction. 
The upper (vacuum side) atoms align alternatively, 
like S$_1$ and S$'_1$, with the same height,
and the lower (bulk side) atoms align alternatively, 
like S$_2$ and S$'_2$, with the same height,
The z coordinates of S$_1$(S$'_1$) and S$_2$(S$'_2$) 
are denoted by $z_1$ and $z_2$, respectively. 
The atoms of S$_{\rm 3a}$ and S$_{\rm 3b}$ are located at the same height
with a  z coordinate denoted by  $z_3$.


The present test problem is to determine
the $N=3$ valuables $z_1$, $z_2$ and $z_3$.
In other words,
the structure data $X$ lies in 
the three-dimensional data space $(X \equiv (z_1, z_2, z_3))$.
Here a numerically generated \lq reference' data $\bm{D}_{\rm ref}$ is used, 
instead of the real experimental data $\bm{D}_{\rm exp}$.  
The reference data $\bm{D}_{\rm ref}$ is generated numerically 
corresponding to the known structure $(z_1, z_2, z_3) = (z_1^{\rm (ref)}, z_2^{\rm (ref)}, z_3^{\rm (ref)})$  
($\bm{D}_{\rm ref} \equiv \bm{D}_{\rm cal}(z_1^{\rm (ref)}, z_2^{\rm (ref)}, z_3^{\rm (ref)}$)
\cite{SHIRASAWA2006}.
The R-factor is defined as
\begin{eqnarray}
R(z_1, z_2, z_3) \equiv \left| \bm{D}_{\rm cal}(z_1, z_2, z_3) - \bm{D}_{\rm cal}(z_1^{\rm (ref)}, z_2^{\rm (ref)}, z_3^{\rm (ref)}) \right|.
\label{EQ-R-FACTOR-DEM}
\end{eqnarray}
The reference data is set to be $(z_1^{\rm (ref)}, z_2^{\rm (ref)}, z_3^{\rm (ref)}) 
\approx (5.231 \AA, 4.371 \AA, 3.596 \AA)$.  
The precision on the order of $10^{-3}$ \AA \, is beyond experimental spatial resolution
but is left here,
because the present numerical test is carried out 
to demonstrate that the present two-stage analysis method can reach the minimum point
with a numerical precision finer than experimental spatial resolution.

Figure ~\ref{FIG-SURFACE}(c) shows
the real part of the scattering potential $({\rm Re}[U_0(z)])$ 
in the case of the reference structure averaged over the $x$-$y$ plane.  
The peaks of the function $({\rm Re}[U_0(z)])$ are located 
at the atomic positions, such as $z=z_1^{\rm (ref)}, z_2^{\rm (ref)}$ and $z_3^{\rm (ref)}$.

The present numerical computation was carried out 
on the supercomputer \lq Sekirei' at the the Institute for Solid State Physics, the University of Tokyo.
One processor node consists of two Intel Xeon 2.5 GHz (12 core) processors.

\begin{figure}[h]
\begin{center}
  \includegraphics[width=0.5\textwidth]{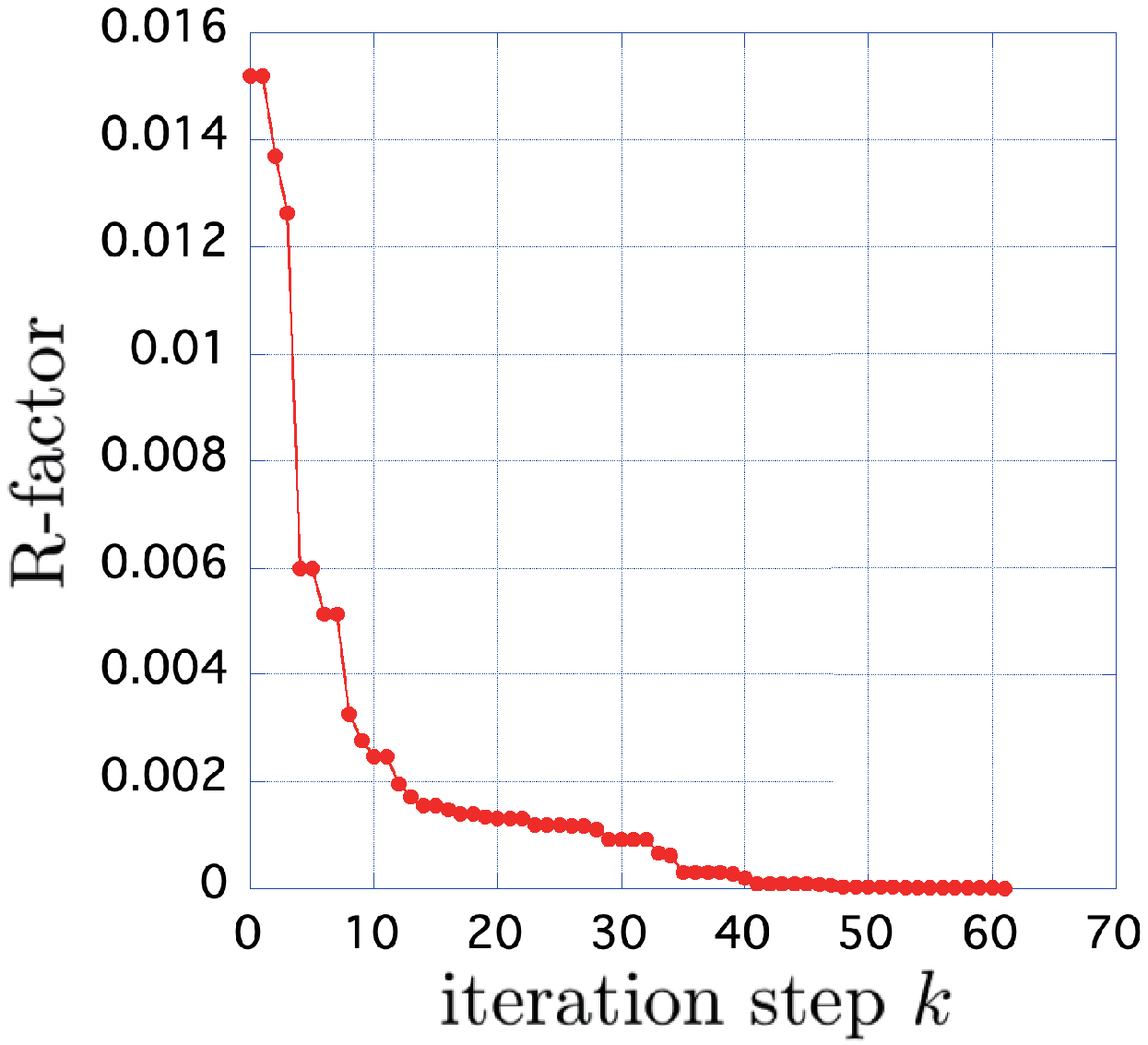}
\end{center}
\caption{
The iterative refinement of the R-factor in the local search.
}
\label{FIG-SEARCH-2}       
\end{figure}

\subsection{Result}

In the first stage, 
the grid points are generated with an equi-interval of $h=0.25 \AA$ in
the region of $z_{\rm min} \equiv 1.00 \AA \le z \le z_{\rm max} \equiv 6.00 \AA $.
The number of the prepared grid points for $(z_1, z_2, z_3)$
is $M' = (((z_{\rm max}-z_{\rm min})/h)+1)^3 =21^3=9,261$.
In addition, 
the constraint of $z_3 \le z_2 \le z_1$ is imposed on the grid points
and the number of grid points is reduced to $M=1,771$.
The global search is carried out with $P_{\rm node}=$8 processor nodes or 
$P_{\rm core} = 8 \times 2 \times 12 = 192$ processor cores. 
Each processor core performs one MPI process and is responsible for
the computation on $M/P_{\rm core} = 1,771/ 192 \approx 9$ grid points.
The total elapsed time for the global search is $T_{\rm global} \equiv 13$ s,
where the elapsed time includes the computational time and the file I/O time.
As results,
the first, second and third optimal grid points are found at
$X^{\rm (g1)} \equiv (z_1^{(g1)}, z_2^{(g1)}, z_3^{(g1)}) = (5.50 \AA, 4.75 \AA, 3.75 \AA) $,
$X^{\rm (g2)} \equiv (z_1^{(g2)}, z_2^{(g2)}, z_3^{(g2)}) = (5.25 \AA, 4.25 \AA, 3.50 \AA) $,
and 
$X^{\rm (g3)} \equiv (z_1^{(g3)}, z_2^{(g3)}, z_3^{(g3)}) = (5.25 \AA, 4.75 \AA, 3.75 \AA) $,
respectively.
Their R-factor values are 
$R(X^{\rm (g1)}) \approx 0.0150$, 
$R(X^{\rm (g2)}) \approx 0.0152$, 
$R(X^{\rm (g3)}) \approx 0.0193$. 


In the second stage,
the local search algorithm was carried out with a single node. 
The methodological details are described
in the previous paper \cite{TANAKA2020_ACTA_PHYS_POLO}. 
The local search algorithm is an iterative refinement procedure and
the first, second and third optimal grid points,
$X^{\rm (g1)}$, $X^{\rm (g2)}$ and $X^{\rm (g3)}$,
are chosen to be the initial points.
(i) When the initial point is chosen to be $X^{(0)}=X^{\rm (g1)}$,
the iterative refinement of the R-factor was converged to $R \approx 0.0058$ after $k=50$ iterative steps and
the converged point is 
$X^{\rm (sol,1)} = (z_1^{\rm (sol,1)}, z_2^{\rm (sol,1)}, z_3^{\rm (sol,1)}) \approx (5.340 \AA, 4.646 \AA, 3.685 \AA) $.
The difference of the converged point and the reference point
is $\delta X^{\rm (sol,1)} \equiv X^{\rm (sol,1)} - X^{\rm (ref)} \approx (0.11 \AA, 0.27 \AA, 0.09 \AA) $.
The total elapsed time for the local search is $T_{\rm local} \equiv 48$ s.
(ii) When the initial point is chosen as $X^{(0)}=X^{\rm (g2)}$, 
the iterative refinement of the R-factor was converged to 
$R=7 \times 10^{-6}$ after $k=61$ iterative steps and
the converged point is 
$X^{\rm (sol,2)} = (z_1^{\rm (sol,2)}, z_2^{\rm (sol,2)}, z_3^{\rm (sol,2)}) \approx (5.231 \AA, 4.371 \AA, 3.596 \AA) $.
The difference between the converged and reference points,
$\delta X^{\rm (sol,2)} \equiv X^{\rm (sol,2)} - X^{\rm (ref)}$,
is of the order of $10^{-5}$ \AA.
The total elapsed time for the local search is $T_{\rm local} \equiv 15$ s.
(iii) When the initial point is chosen to be $X^{(0)}=X^{\rm (g3)}$,
the iterative refinement of the R-factor was converged to 
$R \approx 0.0058$ after $k=47$ iterative steps and
the converged point is 
$X^{\rm (sol,3)} = (z_1^{\rm (sol,3)}, z_2^{\rm (sol,3)}, z_3^{\rm (sol,1)}) \approx (5.340 \AA, 4.645 \AA, 3.684 \AA)$.
The difference between $X^{\rm (sol,3)}$ and $X^{\rm (sol,1)}$ 
is of the order of $10^{-3}$ \AA.
In conclusion, we obtained the true minimum $X^{\rm (sol,2)}$ 
and a local minimum $X^{\rm (sol,1)} (\approx X^{\rm (sol,3)})$.

\begin{figure}[h]
\begin{center}
  \includegraphics[width=0.5\textwidth]{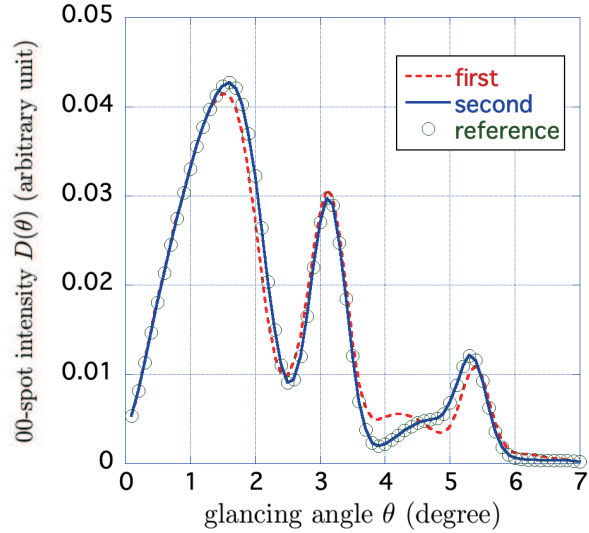}
\end{center}
\caption{
The resultant rocking curves in the first and second stages
are plotted as the dashed and solid lines, respectively.
The reference rocking curve is also plotted as circles.
}
\label{FIG-SEARCH-3}       
\end{figure}

Figure \ref{FIG-SEARCH-1} shows 
partial regions of the global search. 
The calculated value of the R-factor $R=R(z_1, z_2, z_3)$ 
is plotted at each grid point as colored circles on the plane of $z_3=z_3^{(g2)}=3.50 \AA$ 
in Figure \ref{FIG-SEARCH-1}(a)
and on the plane of $z_1=z_1^{(g2)}=5.25 \AA$
in Figure \ref{FIG-SEARCH-1}(b).
One can find that
the R-factor is sensitive not only to the top-most surface atoms
$(z_1,z_2)$ but also to the atoms below $(z_3)$. 
Figure ~\ref{FIG-SEARCH-2} shows
the iterative refinement of the R-factor with the initial point of $X^{(0)}=X^{\rm (g2)}$.
Figure ~\ref{FIG-SEARCH-3} shows
the rocking curves for $X=X^{\rm (g2)}$ obtained in the first stage
and for $X=X^{\rm (sol,2)}$ obtained in the second stage.
The rocking curve is plotted also for the reference point $X=X^{\rm (ref)}$. 

Here several points are noted: 
(i) the present computation demonstrates that 
the two-stage algorithm reaches the exact solution ($R \rightarrow 0$)
for a numerically generated reference data. 
When one analyzes real experimental data, 
an optimized value of $R \le 10^{-2}$ is usually acceptable,
owing to the uncertainties involved in the experimental data;
(ii) the second stage can be parallelized, as well as the first one,
since one can run multiple local search processes with different initial points, $X^{(0)}$. 
Possible candidates of the initial points are
(local) minima or saddle points on grid.

Finally, we should comment on the potential limitation
of the grid-based global search.
The number of grid points $M$ is usually proportional to $m^N$
($M \propto m^N$),
where $N$ is the number of the variables and
$m$ is the number of grid points on one axis. 
Consequently,
the grid-based global search will incur a huge computational cost,
when the number of variables $N$ is larger $(N \gg 3)$. 
Such a case should be analyzed by 
Bayesian inference with a Monte Carlo algorithm. 
The Monte Carlo algorithm is known as a reliable and efficient sampling method.
The method provides the posterior probability of atomic positions through Bayes' theorem 
and enables us to evaluate the uncertainty of estimated atomic positions. 
This method has been applied to surface structure analysis 
by X-ray diffraction investigations\cite{ANADA2017, ANADA2018}. 
We are now developing another global search method
with a massively parallel Monte Carlo algorithm
\cite{HUKUSHIMA_PAMC}.

\section{Summary \label{SEC-SUMMARY} } 

The present paper proposes a two-stage analysis method
for the surface structure determination
by total-reflection high-energy positron diffraction (TRHEPD) experiments.
The algorithm realizes a global search that does not requires any initial guess. 
A test problem is solved with a numerically generated reference data.
The software is  based on the inverse problem,
where the forward problem is a quantum scattering problem or partial differential equation.
Analysis of real experimental data is ongoing. 
The program code will be available online in the near future.

A future aspect of this research will be the extension of this method into a more general data analysis software on surface structure, not only for positron diffraction but also for X-ray and electron diffraction by substitution of the forward problem solver.

\jwakg{}

The present research is supported by the Japanese government in the post-K project and KAKENHI funds (19H04125, 17H02828). 
Numerical computations were also carried out at the facilities of the Supercomputer Center, the Institute for Solid State Physics, the University of Tokyo 
Several numerical computations were also carried out 
by the supercomputer Oakforest-PACS for 
(i) Initiative on Promotion of Supercomputing for Young or Women Researchers, Information Technology Center,
The University of Tokyo, (ii) the HPCI project (hp190066) , (iii) Interdisciplinary Computational Science Program in the Center for Computational Sciences, University of Tsukuba. The supercomputer at the Academic Center for Computing and Media Studies, Kyoto University was also used
for the program development.

\end{document}